\documentclass[prb,amsmath,amssymb]{revtex4}
\pdfoutput=1
\usepackage{graphicx}
\usepackage{dcolumn}
\usepackage{bm}
\usepackage{amsmath}
\usepackage{color}
\usepackage{units}
\usepackage{booktabs}
\usepackage{ae} 

\DeclareFontFamily{U}{euc}{}
\DeclareFontShape{U}{euc}{m}{n}{<-6>eurm5<6-8>eurm7<8->eurm10}{}%
\DeclareSymbolFont{AMSc}{U}{euc}{m}{n} 
\DeclareMathSymbol{\umu}{\mathord}{AMSc}{"16}

\newcommand{\beq} {\begin{equation}}
\newcommand{\eeq} {\end{equation}}

\begin{document}


\title{An on-chip diamond optical parametric oscillator}

\author{B. J. M. Hausmann$^1$*, I. Bulu$^1$*, V. Venkataraman$^1$*, P. Deotare$^2$ and M. Lon\v{c}ar$^1$}
\affiliation{1. Harvard University, School of Engineering and Applied Sciences, Cambridge, MA, USA}
\affiliation{2. Massachusetts Institute of Technology, Research Laboratory of Electronics, Cambridge, MA, USA}

\maketitle

\textbf{Efficient, on-chip optical nonlinear processes are of great interest for the development of compact, robust, low-power consuming systems for applications in spectroscopy, metrology, sensing and optical and quantum optical information processing \cite{Udem2002, Kippenberg2011}. Diamond holds promise for these applications, owing to its exceptional properties \cite{Aharonovich2011, Zaitsev2001}. However, although significant progress has been made in the development of an integrated diamond photonics platform \cite{Faraon2011, Hausmann2012}, optical nonlinearities in diamond have not been explored much apart from Raman processes in bulk samples \cite{Mildren2008}. Here, we demonstrate optical parametric oscillations (OPO) via four wave mixing (FWM) in single crystal diamond (SCD) optical networks on-chip consisting of waveguide-coupled microring resonators. Threshold powers as low as $\unit[20]{mW}$ are enabled by ultra-high quality factor ($\sim1\cdot10^6$) diamond ring resonators operating at telecom wavelengths, and up to 20 new wavelengths are generated from a single-frequency pump laser. We also report the inferred nonlinear refractive index due to the third-order nonlinearity in diamond at telecom wavelengths.} 

To date, on-chip nonlinear nanophotonic systems have been realized in various platforms, including silica \cite{Ferrera2008}, silicon \cite{Turner2008}, $\mathrm{Si_3N_4}$ \cite{Levy2009}, and III-V materials \cite{Hartl2005}. Diamond, which has recently emerged as a promising platform for on-chip photonics \cite{Hausmann2013}, intrinsically combines the advantages of a high refractive index (n=2.4), low absorption losses within its large transmission window (from UV to mid-IR). In addition, a relatively high nonlinear refractive index \cite{Levenson1974, Boyd2008} ($n_2=\unit[1.3\cdot10^{-19}]{m^2/W}$ for visible wavelengths) and the lack of two-photon absorption (owing to its large bandgap) make diamond an attractive candidate for on-chip nonlinear optics in the visible - infrared wavelength range. Furthermore, diamond also offers excellent thermal properties (high thermal conductivity and low thermo-optic coefficient) which allows for high power handling capabilities \cite{Nebel2004}. One interesting application of a nonlinear diamond photonics platform lies in the field of quantum optics:  diamond nonlinearities could allow for frequency translation (to the telecom wavelength range for example) and pulse shaping \cite{Lavoie2013, Raymer2012} of single photons generated by its numerous color centers (often emitting in the visible). These processes promise to combine quantum information science with classical optical information processing systems on the same chip. Another exciting  application of an integrated nonlinear diamond photonics platform is the realization of microresonator based high-repetition rate frequency combs \cite{Kippenberg2011}. On-chip frequency combs, with repetition rates reaching the THz range, have recently been realized via the nonlinear Kerr effect in high quality (Q) factor microresonators fabricated in a number of different materials including silica and silicon nitride \cite{Del'Haye2007, Razzari2009, Levy2009, Okawachi2011}. Diamond's unique properties mentioned above potentially allows for the realization of temperature-stabilized frequency combs operating over wide wavelength range \emph{on the same chip}.

Owing to an inversion symmetry in crystal lattice, diamond's lowest order nonzero nonlinear susceptibility \cite{Boyd2008} is $\chi^{(3)}$. A third-order nonlinear parametric process where two pump photons at frequency $\nu_P$ are converted to two different photons at $\nu_+$ and $\nu_-$  (denoted as signal and idler, respectively), such that energy conservation is satisfied by $2\nu_p=\nu_+ + \nu_-$ is called four-wave mixing (FWM). The FWM gain scales with the pump intensity, and the pump power requirement can be reduced by confining the light to nano-waveguides \cite{Foster2006}. 
In addition to energy conservation, FWM in a waveguide also entails momentum conservation, or phase-matching which implies $\Delta k = 2\gamma P_{p}-\Delta k_L \sim 0$ \cite{Foster2006, Hansryd2002}. Here, the second term $\Delta k_L = 2k_{p} - k_+ - k_-$ is the phase mismatch due to the linear dispersion ($k_{p}$, $k_+$ and $k_-$ are the pump, signal and idler wavenumbers, respectively), $\gamma=2\pi\nu_{p}n_2/cA_{eff}$ is the effective nonlinearity and $A_{eff}$ the effective optical mode area. The term $2\gamma P_{p}$ arises from the nonlinear response to the strong pump, which imposes self-phase modulation (SPM) on itself and cross-phase modulation (XPM) on the generated modes that is twice as large as the SPM \cite{Del'Haye2007, Kippenberg2004}. This nonlinear phase shift needs to be compensated for by the linear dispersion, i.e. $\Delta k_L > 0$. Consequently, the group velocity dispersion (GVD) of the optical mode needs to be anomalous around the pump wavelength \cite{Foster2006, Hansryd2002}, that is GVD $=-\frac{\lambda}{c} \cdot d^2n_{eff}/d\lambda^2 > 0$, where $n_{eff}$ is the effective index of the waveguide mode, $\lambda$ is the wavelength and $c$ is the speed of light in vacuum. 

\begin{figure}[htbp]
\centering\includegraphics[width=\textwidth]{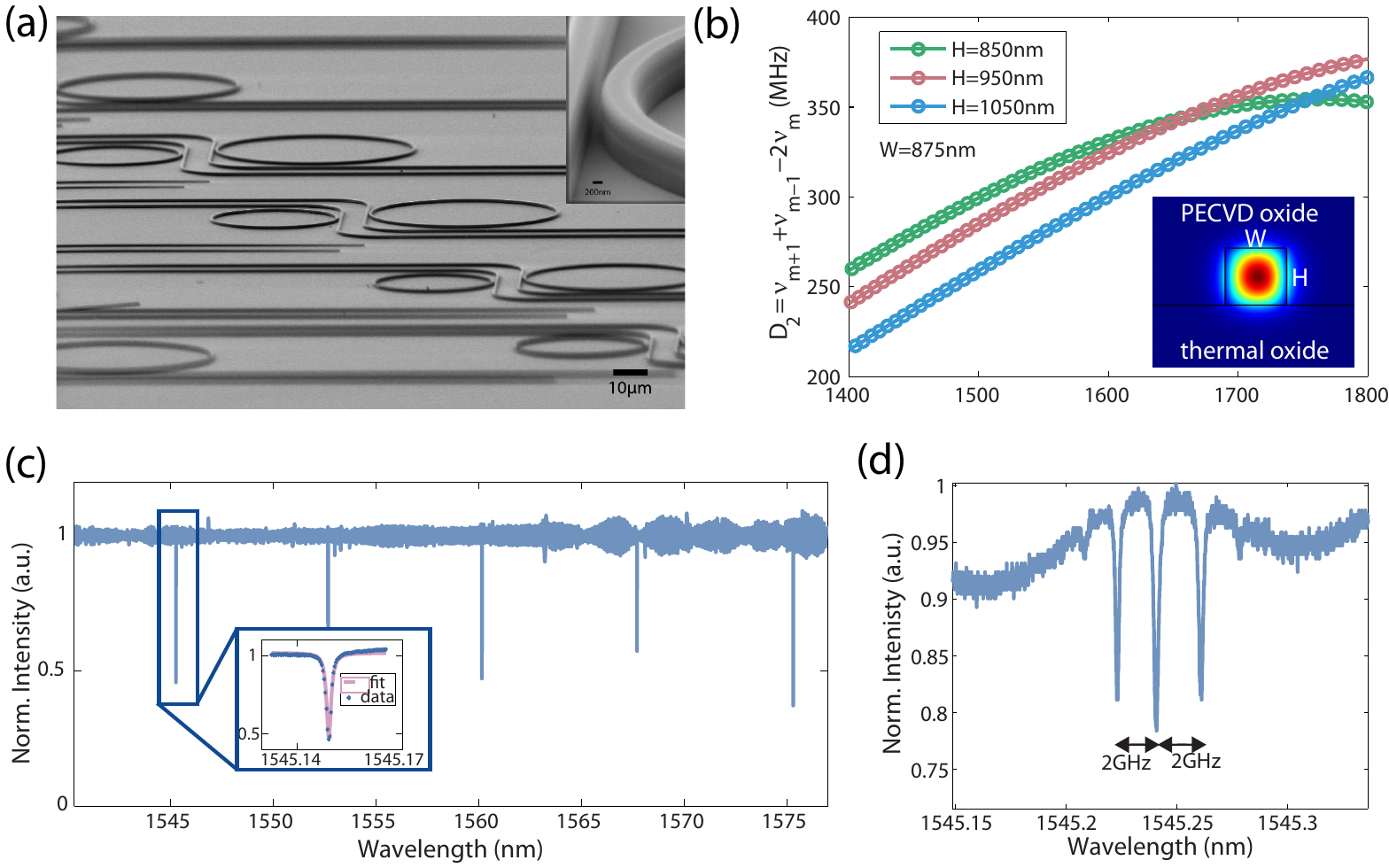} 
\caption{\textbf{Integrated ultra-high Q single-crystal diamond (SCD) ring resonators.} \textbf{(a)} Scanning electron microscopy (SEM) image of an array of waveguide-coupled SCD ring resonators on a $\mathrm{SiO_2/Si}$ substrate. Before testing, chips are covered with $\unit[3]{\mu m}$ of PECVD deposited silica. The inset shows a magnified view of the ring - waveguide coupling section with a $\unit[\sim475]{nm}$ gap size for the measured device. The rings are $\unit[\sim850]{nm}$ high, $\unit[\sim875]{nm}$ wide and have $\unit[20-30]{\mu m}$ radii. \textbf{(b)} Robust dispersion engineering for a ring width of $\unit[875]{nm}$ allows for a range of ring heights to yield anomalous dispersion in the wavelength range of interest. The inset shows the optical ring resonator mode profile in the diamond waveguide surrounded by silica. \textbf{(c)} Normalized transmission spectrum of a ring resonator reveals high $Q$-factor modes. The radius of the ring is $\unit[20]{\mu m}$, corresponding to a free-spectral range (FSR) of $\sim\unit[7.5]{nm}$ ($\sim$925 GHz). Inset: A loaded $Q$-factor of $Q_L \sim 1\cdot10^6$ is inferred from a Lorentzian fit for the mode at $\unit[1545.1]{nm}$. \textbf{(d)} The light from our tunable laser is phase-modulated at $\unit[2]{GHz}$ to produce sidebands that are then used as a ruler to calibrate the wavelength axis in our transmission measurements. Using this approach, loaded  $Q$ factors as high as $Q_L \simeq 1.14\cdot10^6$ were estimated. 
}
\label{fig1}
\end{figure}

The FWM efficiency can be drastically increased by using high-$Q$ resonators \cite{Absil2000, Turner2008}, where photons make multiple round trips on resonance, resulting in the optical intensity to be enhanced by a factor of the finesse. Optical parametric oscillation is achieved when the round trip FWM gain exceeds the loss in the resonator, a process analogous to a laser above threshold, and bright coherent light is generated at the signal and idler wavelengths. In our diamond ring resonators (Figure \ref{fig1}) momentum is intrinsically conserved since the optical modes are angular momentum eigenstates \cite{Vahala2003}. In this case, anomalous dispersion is required to meet energy conservation between the cavity modes with different angular momentum ($m$) that participate in a FWM process \cite{Del'Haye2007}. This implies that the frequency separation between adjacent modes of the ring resonator, $\vert\nu_m - \nu_{m-1}\vert$ (or the free-spectral range, FSR), increases as a function of the mode number $m$. The resonator dispersion $D_2$, given by the change in FSR ($ \nu_{m+1} + \nu_{m-1} - 2\nu_{m}$), thus needs to be positive for modes around the pump wavelength \cite{Herr2012, Del'Haye2007}. The unequal frequency spacing of the resonator modes due to anomalous dispersion is compensated by nonlinear optical mode "pulling", i.e a shift in the resonance frequencies caused by SPM and XPM due to the pump  \cite{Kippenberg2004, Del'Haye2007}.

The intrinsic material dispersion of diamond is normal at telecom wavelengths. The net anomalous waveguide dispersion can be engineered to be anomalous through geometrical dispersion by appropriately designing the cross-sectional dimensions \cite{Foster2006, Levy2009, Okawachi2011, Herr2012} of the waveguide. However, our fabrication technique (see methods) relies on thin SCD films that are typically wedged, resulting in a thickness variation of at least $\unit[>300]{nm/mm}$ across the sample \cite{Hausmann2013}. This effect occurs due to the mechanical polishing process for thin diamond plates ($\unit[\sim20]{\mu m}$ thick) that are used to realize our diamond-on-insulator platform \cite{Hausmann2012}. Therefore, the ring resonator design has to be robust and the dispersion insensitive to the variations in the diamond film thickness. The inset of figure \ref{fig1}b depicts the ring resonator mode profile in our geometry, a diamond ring resonator on top of a $\mathrm{SiO_2/Si}$ substrate and capped with a deposited $\mathrm{SiO_2}$ layer.  Figure \ref{fig1}b shows that in the case of a ring width of $\unit[875]{nm}$, the resonator dispersion can be made anomalous in the wavelength range of interest for a range of film thicknesses (ring heights). Furthermore, for a ring resonator of radius $\unit[20]{\mu m}$, anomalous dispersion for the transverse-electric (TE) mode can be achieved in the $\unit[1300 - 1800]{nm}$ wavelength range for widths $\unit[800 - 900]{nm}$ and heights $\unit[500 - 1000]{nm}$. This is well within our fabrication tolerances - Figure \ref{fig1}a shows waveguide coupled SCD  ring resonators of radii 20- and $\unit[30]{\mu m}$, fabricated according to a method that we have recently presented \cite{Hausmann2013} (see methods). 

To characterize the diamond resonators, we use a fiber-coupled transmission setup that has been described elsewhere \cite{Deotare2012,Hausmann2013}. First, transmission measurements are performed by sweeping a continuous - wave laser (Santec TSL-510) across the telecom wavelength range to measure the resonator $Q$-factors and the coupling of the bus - waveguide to the rings (Fig. \ref{fig1}c). Most devices are found to be slightly under-coupled. Loaded $Q$-factors ($Q_L$) as high as $1\cdot10^6$ are measured for the TE mode, with most devices having $Q_L > 2\cdot10^5$. To ensure an accurate resonance linewidth measurement, a radio-frequency (RF) phase modulation is imparted on the input light which generates sidebands around the main resonance (Fig. \ref{fig1}d). Comparing the linewidth of the resonance with the separation between the side-bands ($\unit[1-3]{GHz}$) allows for a precise calibration of the wavelength/frequency axis \cite{Collot1993}. Using this method, we measure the highest $Q$ factor to be $\sim 1.14\cdot10^6$. 

\begin{figure}[htbp]
\centering\includegraphics[width=0.7\textwidth]{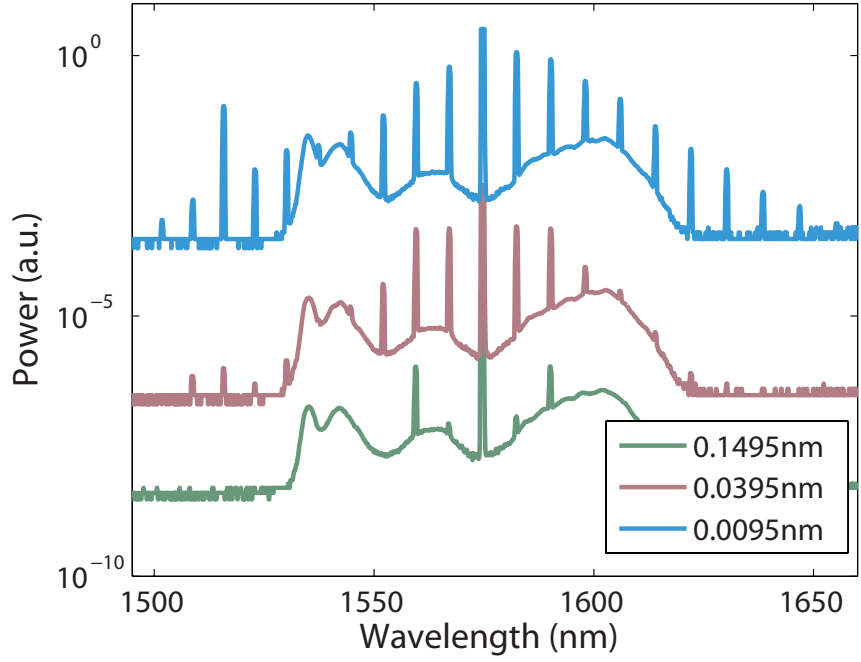} 
\caption{\textbf{Optical parametric oscillation spectrum as a function of blue-detuning from resonance minimum wavelength.} New frequencies are generated in the spectrum as the pump wavelength approaches the resonance (transmission minimum) starting from a blue detuned position. The spectra are logarithmically offset for clarity.
}
\label{fig2}
\end{figure}

High pump powers required for OPO is obtained by sending the input laser through an erbium-doped fiber amplifier (EDFA, Manlight). The pump is initially slightly blue detuned and then slowly moved into resonance. The power absorbed by the ring causes a thermal red-shift of the resonance potentially arising due to the heating of the silica cladding or surface effects at the diamond silica interface. While tuning the laser deeper into resonance, the output light is monitored on an optical spectrum analyzer (OSA, HP 70952B). As the offset of the pump to the resonance minimum decreases, more power is transferred to the ring resonator, eventually resulting in the generation of pairs of new lines - at integer multiples of the resonator FSR - around the pump. The first sidebands are generated at  mode numbers $m \simeq \sqrt{\kappa/D_2\cdot(\sqrt{P_{in}/P_{th} - 1}+1)}$ away from the pump \cite{Herr2012}, where $\kappa$ represents the resonator linewidth (cavity decay rate), $D_2$ is the resonator dispersion discussed earlier, $P_{in}$ is the input pump power and $P_{th}$ is the threshold pump power for parametric oscillation. Tuning the pump deeper into resonance generates several new modes further away from the pump, finally resulting in a spectrum of multiple lines with a frequency spacing given by the FSR (Fig. \ref{fig2}). The pump power coupled into the resonator is intrinsically stabilized during this entire process achieving a thermal 'soft-lock' \cite{DelHaye2008}, and stable oscillation is observed for up to $\unit[\sim20]{min}$ (limited by the fiber-stage drifts). 

\begin{figure}[htbp]
\centering\includegraphics[width=1\textwidth]{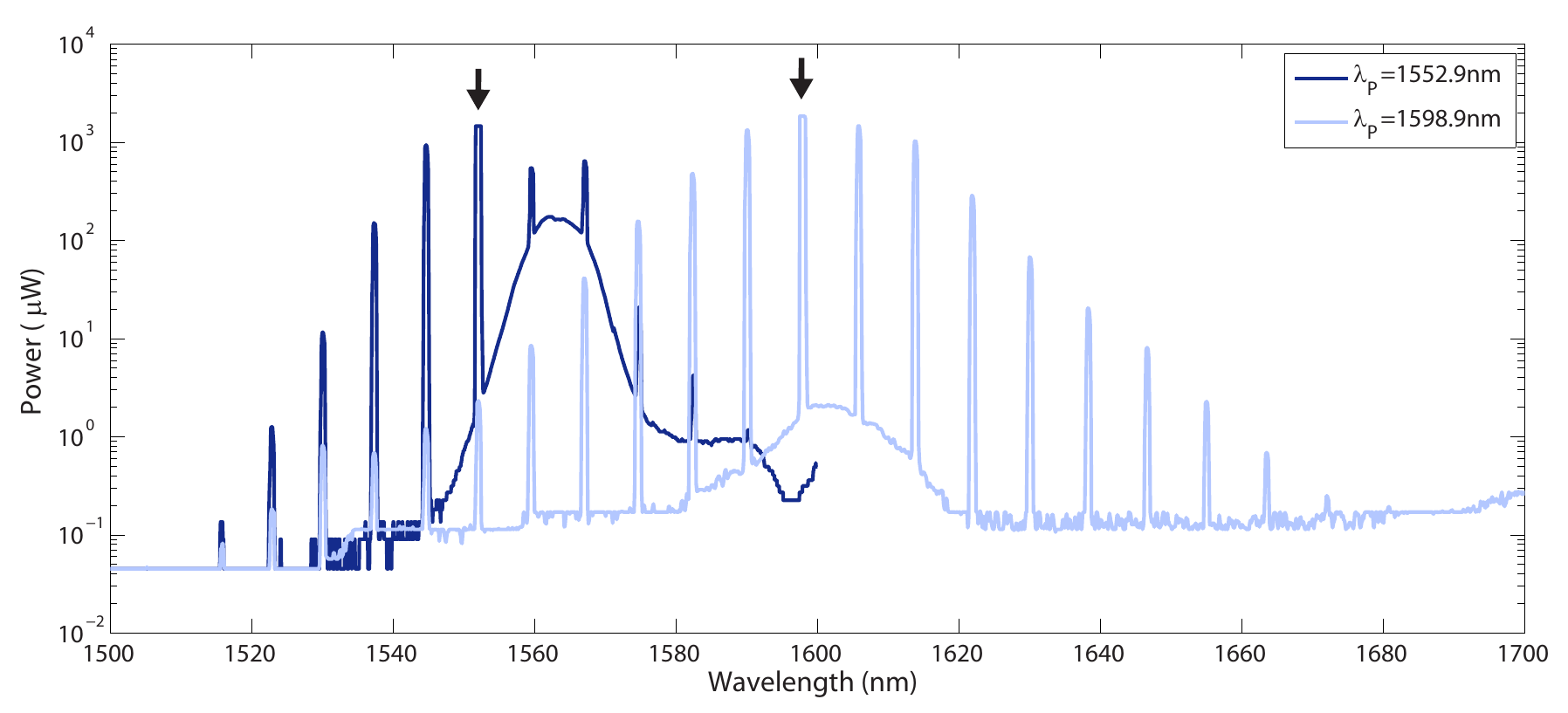} 
\caption{\textbf{Optical parametric oscillation spectra for different pump wavelengths.} The spectra, generated from the same ring resonator for two different pump wavelengths, $\unit[\sim1553]{nm}$ and $\unit[\sim1599]{nm}$ are shown. The pump power is the same in each case ($\unit[\sim80]{mW}$ in the waveguide). A total of 20 new lines are generated when pumping at $\unit[1599]{nm}$, that is 10 lines when pumping at $\unit[\sim1553]{nm}$. This can be explained by a higher coupling efficiency between the bus-waveguide and the ring resonator as well as a stronger dispersion for longer wavelengths.
}
\label{fig3}
\end{figure}

The performance of our diamond OPO device is studied as a function of pump wavelength. The same ring is pumped at two different resonances, first at $\sim \unit[1552]{nm}$ (C-band) and then at $\sim \unit[1598]{nm}$ (L-band), and their output spectra compared (Fig. \ref{fig3}). For the same pump power of $\sim \unit[80]{mW}$ in the waveguide, the former generates 10 new lines spanning a range of $\unit[75]{nm}$ while the latter generates 20 lines spanning a range of $\unit[165]{nm}$. This effect can be explained by an increased power drop into the ring for a larger ring - waveguide coupling efficiency that exists at longer wavelengths in our case (since the rings are under coupled). Additionally, this effect might be associated with the change in dispersion with wavelength.

\begin{figure}[htbp]
\centering\includegraphics[width=1\textwidth]{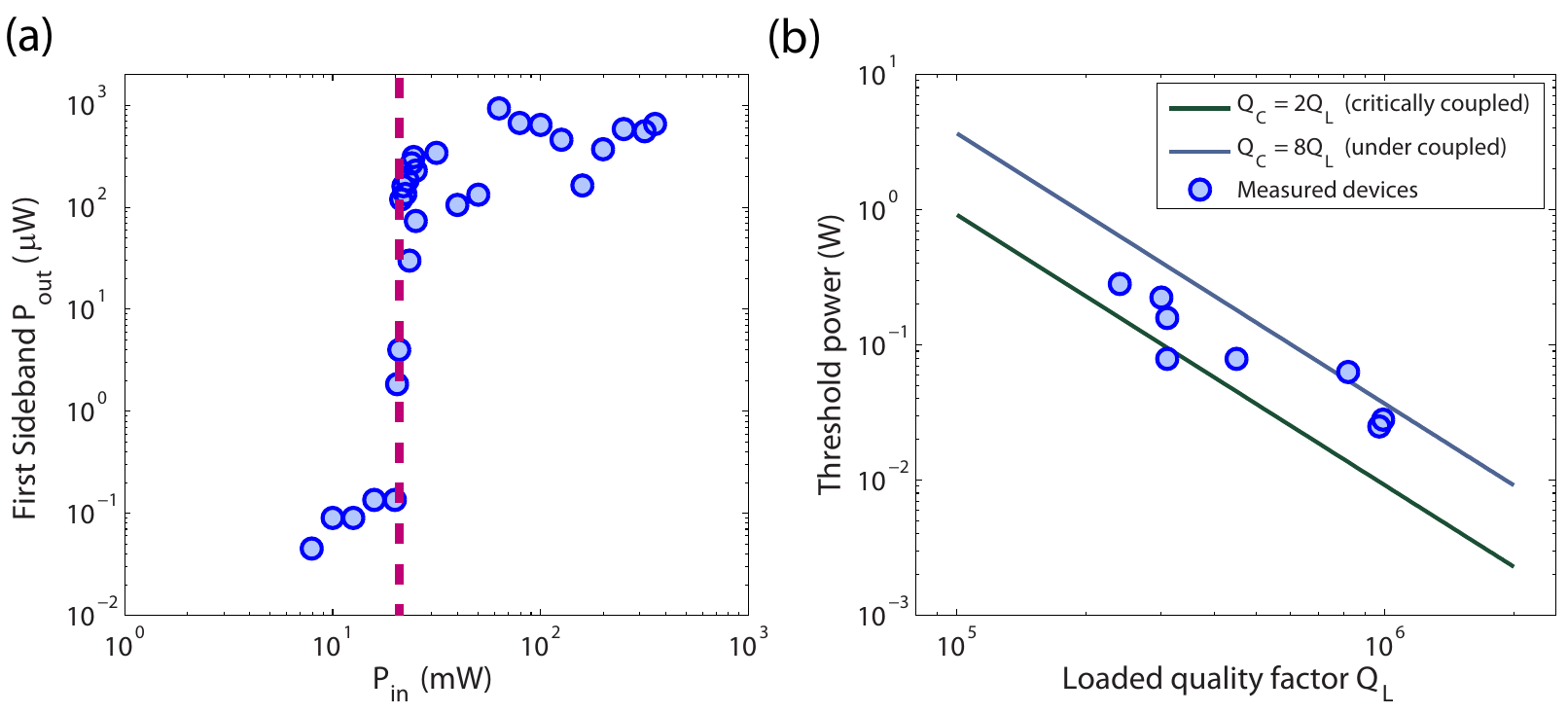} 
\caption{\textbf{Parametric oscillation threshold and its dependence on $Q$-factor.} \textbf{(a)} Output power in the first generated sideband as a function of input pump power (both estimated in the waveguide) for a device with $Q_L = 9.7\cdot10^5$. The threshold for oscillation is observed to be $\unit[20]{mW}$. \textbf{(b)} Threshold powers for oscillation as a function of loaded quality factor $Q_L$ (power estimated in the waveguide). The data is measured for eight different devices (blue dots). Threshold powers roughly follow the theoretically predicted trend of being inversely proportional to $Q_L^2$, with most devices being slightly under-coupled i.e. $Q_C > 2Q_L$ (consistent with the transmission measurements). The black line denotes critical coupling (100\% transmission dip on-resonance), while the blue line denotes under-coupled resonators (50\% transmission dip on-resonance).}
\label{fig4}
\end{figure}

To determine the threshold for parametric oscillation, the output power in the first generated sideband is measured as a function of pump power. Fig. \ref{fig4}a shows the data for a device pumped at a resonance near $\unit[1575]{nm}$ with $Q_L = 9.7\cdot10^5$, where we infer a $P_{th}$ of only $\sim \unit[20]{mW}$ in the waveguide and a conversion slope efficiency of $\sim$2\%. For pump powers above threshold, oscillation occurs into multiple new modes limiting the power converted to the first sideband. We infer the total power in a total of 20 generated modes combined to be $\unit[3.4] {mW}$ (as estimated in the waveguide) for an input pump power of $\unit[316]{mW}$ (in the waveguide) and hence an overall conversion efficiency of $\sim1.1$\%.   

The threshold power ($P_{th}$) for parametric oscillation arising from the third-order nonlinearity (FWM) can be also estimated from theory as \cite{Matsko2005}:
\begin{equation}
P_{th}\simeq1.54\left(\frac{\pi}{2}\right)\frac{Q_C}{2Q_L}\cdot\frac{n_0^2 V}{n_2 \lambda_P Q_L^2} 
\end{equation}
where $\lambda_P$ is the pump wavelength, $V$ is the resonator mode volume and $n_0$ is the linear refractive index ($\sim$2.4 for diamond). By measuring $P_{th}$ for various devices with different $Q$-factors, the nonlinear refractive index $n_2$ can be inferred in the wavelength range around the pump. The measured $P_{th}$ - estimated in the waveguide- for eight different devices on the same chip is depicted in figure \ref{fig4}b. From this data, we extract the nonlinear refractive index of diamond in the telecom range to be $n_2=\unit[(8.2 \pm 3.5) \cdot10^{-20}]{m^2/W}$ which is about a factor of 1.5 smaller than the $n_2$ value reported for visible wavelengths \cite{Boyd2008, Levenson1974}. This is in good agreement with the theoretical prediction of the dispersion of the nonlinear susceptibility (longer wavelengths being more off-resonant from the bandgap) \cite{Levenson1974}. Fig. \ref{fig4}b also shows that most of the devices measured are on the under-coupled side, consistent with the expectations form the transmission measurements.
 
In summary, we demonstrate an on-chip OPO operating at telecom wavelengths based on a fully-integrated, monolithic, single-crystal diamond micro-resonator. Despite of the non-standard fabrication approach and significant wedging in our diamond films, we could infer a reasonably high device yield of 30\% (out of 26 devices) in a diamond film of a size of $\unit[570]{\mu m}$ x $\unit[630]{\mu m}$. The OPO leverages the $\chi^{(3)}$-nonlinearity of diamond and experimentally measured $n_2=\unit[(8.2 \pm 3.5) \cdot10^{-20}]{m^2/W}$ of diamond in the telecom wavelength range to realize a FWM gain for sidebands around the pump frequency. Ring resonators with $Q$-factors near one million enable oscillation threshold powers as low as $\unit[20]{mW}$ in the bus-waveguide, and 20 sidebands spanning a wavelength range of $\unit[165]{nm}$ are generated with pump powers less than $\unit[100]{mW}$. The total power generated in all sidebands was up to 1.1\% of the pump power. This is comparable to what has been achieved in other material systems \cite{Levy2009}. The OPO process is currently limited by pump power, possibly also absorption losses, and could benefit from stronger dispersion and higher $Q$ factors in general. For example, coupling higher pump powers into the OPO device should enable the generation of broadband, high-repetition-rate optical frequency combs that are desirable for numerous applications \cite{Del'Haye2007, Herr2012, Foster2011, Okawachi2011, Savchenkov2011, Kippenberg2011}. Furthermore, further dispersion engineering combined with studies of the absorption losses (most likely associated with PECVD oxide deposited on our devices and/or surface states at the diamond/oxide interface) that result in the thermal effects observed in our experiments can result in higher $Q$ factors, thus facilitating comb generation. The large thermal conductivity of diamond along with its small thermo-optic coefficient should enable the realization of temperature insensitive frequency combs capable of handling high powers.  Given diamond's large transparency window from the UV to the IR, in addition to the absence of loss mechanisms like two-photon absorption and free-carrier absorption plaguing other material systems for photons up to $\unit[\sim2.75]{eV}$, this technology can be transferred to other wavelengths where on-chip devices have not yet been realized, e.g. the visible range, which is also intriguing for the generation of Raman lasers at exotic wavelengths. Integrating these devices with sources of non-classical light, such as single-photon emitters based on color-centers in diamond \cite{Faraon2011, Hausmann2012}, provides another compelling route. Our work thus opens up an avenue for research in diamond nonlinear photonics, where all-optical information processing on-chip may be realized at both the classical and quantum level.

\section{Methods}\label{methods}
\textbf{Device fabrication:} The fabrication sequence is based on the recently described approach for integrated single-crystal diamond (SCD) devices \cite{Hausmann2013}. A $\unit[20]{\mu m}$ thick type-Ib SCD slab (Element Six) is cleaned in boiling acids (Nitric, Sulfuric, Perchloric in equal ratio), then thinned to a desired device layer thickness, by an oxygen and $\mathrm{Ar/Cl_2}$ based inductively coupled plasma reactive ion etch (ICP RIE). The diamond film is then transferred to a $\mathrm{SiO_2/Si}$ substrate ($\unit[2]{\mu m}$ thick thermal $\mathrm{SiO_2}$ layer). An etch mask is formed by e-beam lithography (Elionix) using XR-1541-6 and Fox 16 e-beam resist (spin-on-glass, Dow Corning), which is then transferred to the diamond film in a second etch step. Next, polymer in- and out-coupling pads consisting of SU-8 resist with a $\unit[3]{\mu m}$ x $\unit[3]{\mu m}$ cross-section are aligned with respect to the adiabatically tapered diamond waveguides in a second e-beam lithography step to extend the diamond waveguides to the ends of the substrate \cite{Deotare2012}. This step is followed by depositing $\unit[3]{\mu m}$ of silica using plasma enhanced chemical vapor deposition (PECVD) to cap the devices and to allow for controlled cleaving and polishing of the end facets.

\textbf{Modeling:} A finite-element mode solver (COMSOL) is used to simulate the diamond ring resonator dispersion. The material dispersion of both the thermally - grown $\mathrm{SiO_2}$ underneath the diamond devices and the capping $\mathrm{SiO_2}$ deposited via plasma-enhanced chemical vapour deposition (PECVD) are evaluated via ellipsometry measurements and this data is  included in the mode calculations. To optimize the coupling into the ring resonator modes, the gap between the coupling waveguide and the ring resonator is designed by 3D Finite-Difference Time Domain (FDTD) simulations (Lumerical). For the above mentioned cross-sectional dimensions, gaps of $\unit[400 - 500]{nm}$ yield coupling $Q$-factors $Q_C > 5\cdot10^5$.

\section{Acknowledgements}
Devices were fabricated in the Center for Nanoscale Systems (CNS) at Harvard. The authors thank Tobias Kippenberg, Ron Walsworth and Mikhail Lukin for helpful discussions as well as Daniel Twitchen and Matthew Markham from Element Six for support with diamond samples. B.H. gratefully acknowledges support by the Harvard Quantum Optics Center (HQOC). This work was supported in part by the National Science Foundation (ECCS-1202157),  as well as AFOSR MURI (grant FA9550-12-1-0025).  

\section{Additional information} 
Correspondence and requests for materials should be addressed to M.L..

\end{document}